\documentclass[prd,showpacs, amssymb,superscriptaddress,aps,notitlepage]{revtex4}

\usepackage{graphicx}
\usepackage{amssymb}
\usepackage{amsmath}
\usepackage{bm}
\usepackage{soul}

\usepackage{mathrsfs}
\usepackage[colorlinks=true]{hyperref}
\usepackage[all]{hypcap} 

\usepackage{color}
\usepackage{multirow}

\usepackage{palatino}
\input{epsf}

\usepackage{journals}
\begin{document}
%%%%%%%%%%%%%%%%%%%%%%%%%%%%%%%%%%

\title{A Semi-analytic Study of Axial Perturbations of Ultra Compact Stars }
\author{Sebastian H. V\"olkel }
\email{sebastian.voelkel@uni-tuebingen.de}
\affiliation{Theoretical Astrophysics, IAAT, University of T\"ubingen, Germany}
\author{Kostas D. Kokkotas}%\thanks{kostas.kokkotas@uni-tuebingen.de}}
\affiliation{Theoretical Astrophysics, IAAT, University of T\"ubingen, Germany}
\date{\today}
% % % 
%%%%%%%%%%%%%%%%%%%%%%%%%%%%%%%%%%
\begin{abstract}
Compact object perturbations, at linear order,  often lead in solving one or more coupled wave equations. The study of these equations 
was typically done by numerical or semi-analytical methods. The WKB method and the associated Bohr-Sommerfeld rule have been proved  extremely useful tools in the study of black-hole perturbations and the estimation of the related quasi-normal modes. Here we present an extension of the aforementioned semi-analytic methods in the study of perturbations of ultra-compact stars and gravastars. 
\end{abstract}

\pacs{04.40.Dg, 04.30.-w , 04.25.Nx, 03.65.Ge}

% 04.40.Dg	Relativistic stars: structure, stability, and oscillations (see also 97.60.-s Late stages of stellar evolution)
% 04.30.-w	Gravitational waves 
% 04.25.Nx	Post-Newtonian approximation; perturbation theory; related approximations
% 03.65.Ge	Solutions of wave equations: bound states

\maketitle
%%%%%%%%%%%%%%%%%%%%%%%%%%%%%%%%%%
%
%%%%%%%%%%%%%%%%%%%%%%%%%%%%%%%%%%
\section{Introduction}
\label{intro}
The recent stunning detection of gravitational waves from two merging black holes by the LIGO interferometers \cite{LIGO1} opens not only a completely new window to astronomy of extreme astrophysical systems, but also paved the way for new fundamental tests of general relativity and gravitation in general \cite{LIGO2}. Black holes and compact stars are among the most promising sources for future detections and provide both, an unprecedented test of gravity's strong field regime and of matter under extreme conditions. 
A very important part of the gravitational wave spectrum is associated with the ringdown phase and the excitation of the so-called quasi-normal modes (QNMs). They are the characteristic oscillation patterns of compact objects represented by damped sinusoids.  This specific part of the spectrum carries unique information about the nature and the characteristics of the emitting source and it is subject to intense study for nearly six decades. The black hole and neutron star seismology has been the subject of numerous analytic, semi-analytic and numerical studies. The recent detection of the quasi-normal mode in the post merger signal of GW150914 \cite{LIGO1} and GW151226 confirmed the predictions and justified the efforts. %\textcolor{red}{Add references? They only measured one full QNM in the first event?}
Accurate measurements of the quasi-normal mode spectrum may be used to test alternative theories of gravity \cite{LIGO2}.  Still the low SNR ($\rho \sim 7$) left room for alternatives to black-holes \cite{2016PhRvL.116q1101C,2016PhLB..756..350K}. Objects which do not have a horizon can potentially produce similar ringdowns, at least to the accuracy of the current measurements.

The study of gravitational perturbations of compact objects in a general relativistic framework was initiated by the pioneering works of Regge and Wheeler \cite{1957PhRv..108.1063R} and Zerilli \cite{1970PhRvL..24..737Z} for the Schwarzschild black hole. It was later extended to Kerr black holes by Teukolsky \cite{1973ApJ...185..635T}. At the same time Thorne and collaborators initiated the study of neutron star perturbations \cite{1967ApJ...149..591T}. For the aforementioned non-rotating and spherical symmetric cases two different types of perturbations equations were defined depending on their parity, the so-called ``axial'' and ``polar'' perturbations. For black holes, the perturbations are obviously only related to spacetime and they are iso-spectral. In the case of stars, the presence of matter leads to a more involved problem since one has to study coupled systems of wave equations for the fluid and spacetime.  The associated modes characterise the equation of state of the fluid \cite{1996PhRvL..77.4134A} while the part of the spectrum related to the spacetime perturbations is associated to the geometry and in GR a new class of modes emerges, the so-called ``spacetime or $w$-modes'' \cite{1992MNRAS.255..119K}.

Moreover, for non-rotating stars, it can be proved that axial perturbations do not couple to the fluid \cite{1967ApJ...149..591T} and thus their perturbation equations reduced to a single radial one as for the case of black holes. This type of perturbations belong to the class of ``spacetime'' quasi-normal modes and were first studied for constant density stars in \cite{1991RSPSA.432..247C, 1991RSPSA.434..449C, 1994MNRAS.268.1015K,PhysRevD.60.024004}, while further numerical studies have been performed over the last twenty years. 
 Comprehensive reviews on  quasi-normal modes of black holes and compact stars as well as an overview about the different families of modes can be found in  \cite{1999LRR.....2....2K, 1999CQGra..16R.159N, 2009CQGra..26p3001B, 2011RvMP...83..793K}.

Einstein's equations are highly non-linear and thus for many cases when the non-linearities are not important we tend to work with their linearized versions. This means that we assume a small perturbation ${h_{\mu \nu}}$ to the background metric ${\tilde g}_{\mu \nu}$, i.e. 
%
%%%%%%%%%%%%%%%%%%%%%%%%%%%%%%%%%%
$
{g_{\mu \nu}} = {\tilde g}_{\mu \nu} + {h_{\mu \nu}}.
$
%%%%%%%%%%%%%%%%%%%%%%%%%%%%%%%%%%
The linearized Einstein's equations, assuming a proper choice of the gauge, reduce to wave equations for the components of ${h_{\mu \nu}}$.
If the background metric is spherically or spheroidally symmetric one can make use of tensor harmonics to expand ${h_{\mu \nu}}$. For spherically symmetric spacetimes describing black holes and for the axial perturbations of spherically symmetric compact stars, the linearized equations reduce to a single one-dimensional wave equation for the radial component $\Psi(x)$ of the {\tt rr} component of ${h_{\mu \nu}}$ \cite{1967ApJ...149..591T,1991RSPSA.434..449C}
%
%%%%%%%%%%%%%%%%%%%%%%%%%%%%%%%%%%
\begin{equation}
\frac{\mathrm{d}^2}{\mathrm{d}{x}^2} \Psi(x) + \left(\omega_n^2-V(x) \right) \Psi(x) = 0 \, ,
\label{wave-equation}
\end{equation}
%%%%%%%%%%%%%%%%%%%%%%%%%%%%%%%%%%
%
where $V(x)$ is an effective potential. The above equation together with the appropriate boundary conditions at the center/horizon and at infinity constitute a boundary value problem for which $\omega_n$ are the eigenvalues. 
This eigenvalue problem will be the subject of our study in the next sections.
 Here we will present a technique based on the Bohr-Sommerfeld rule, which allows for quick and reliable semi-analytical treatment of ultra-compact stellar perturbations. This technique can be also applied to the more exotic compact objects, such as the so-called gravastars \cite{2001gr.qc.....9035M}. We also propose a purely analytic scheme which can be used for all type of potentials as the one shown in Fig. \ref{fig:1}. Thus without the need of elaborate numerical codes one can test alternatives to black-holes or other compact objects.

This paper is organized as follows. In Sec. \ref{Bohr-Sommerfeld Rule} we introduce the basic ideas related to the Bohr-Sommerfeld rule.
We show its generalizations and we describe the techniques that one may use to get accurate results. In Sec. \ref{analytictoymodel} we derive a purely analytic formula for determining the spectra of the semi-bound states met in this type of problems. In Sec. \ref{applications} we apply the method to ultra-compact uniform density stars and gravastars. In Sec.\ref{Discussion} we summarize and discuss our findings. In Sec. \ref{Appendix} we provide additional material.
Throughout the paper, we assume $G=c=\hbar=2m=1$.

%
%\FloatBarrier
%%%%%%%%%%%%%%%%%%%%%%%%%%%%%%%%%%
\section{The Bohr-Sommerfeld Rule}
\label{Bohr-Sommerfeld Rule}
%%%%%%%%%%%%%%%%%%%%%%%%%%%%%%%%%%
%
The classical Bohr-Sommerfeld (BS) rule  is a well known method to obtain approximate results for the energy spectrum $E_n$ of bound states in a confining potential $V(x)$
\begin{equation}
\label{sec:2 cBS}
\int_{x_0}^{x_1} \sqrt{E_n - V(x)} \mathrm{d} x 
= \pi \left(n+\frac{1}{2} \right).
\end{equation}
Here $x_0$ and $x_1$ are classical turning points defined by the root of the integrand and  depend on the value of  $E_n$. It was introduced in the early times of quantum mechanics but can also be derived as a low order result of WKB theory for the Schr\"odinger equation, independent of any physical motivation \cite{1978amms.book.....B}. In this form it was used for the study of high-overtone normal modes of Schwarzschild black holes  \cite{1990CQGra...7L..47G}. Actually, a more general form of the rule can be found  in \cite{1978amms.book.....B,PhysRev.41.721} and in this form it was possible to be extended to the complex plane for the study of quasi-normal modes of Kerr black holes \cite{1991CQGra...8.2217K} and with the more involved phase integral method for Schwarzschild black holes \cite{0264-9381-10-4-010}.  The BS rule was also used for the study of the QNM spectra of AdS black holes \cite{Festuccia:2009:1936-6612:221}.%
%
%%%%%%%%%%%%%%%%%%%%%%%%%%%%%%%%%%
\subsection{Generalized Bohr-Sommerfeld Rule}
\begin{figure}
	\centering
	\includegraphics[width=9cm]{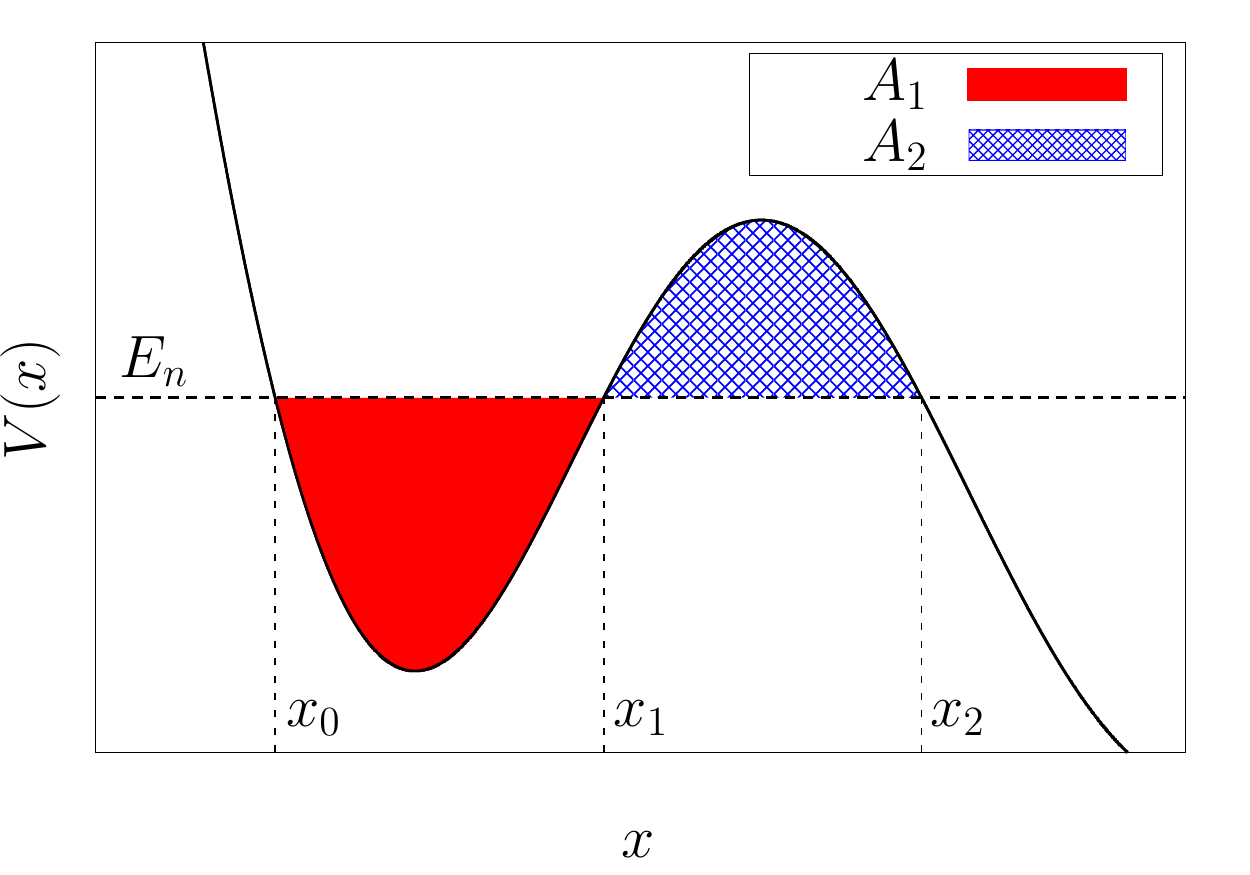}
	\caption{Characteristic potential $V(x)$ for quasi-stationary states. The three classical turning points $x_0$, $x_1$ and  $x_2$ associated to a given energy state $E_n$ are also shown. Qualitatively, the shape of the red region $A_1$ characterises the real part of the energy spectrum, $E_n$, while the imaginary part is characterised by the shape of the blue region $A_2$.}
	\label{fig:1}
\end{figure}

%%%%%%%%%%%%%%%%%%%%%%%%%%%%%%%%%%
%
Using WKB theory it is possible to include higher oder corrections and to generalize the Bohr-Sommerfeld  rule to other type of potentials. This extension has been discussed in \cite{1991PhLA..157..185P} for the case of quasi-stationary states, which can be met in potentials similar to the one drawn in Fig. \ref{fig:1}. 
%
%%%%%%%%%%%%%%%%%%%%%%%%%%%%%%%%%%
%%%%%%%%%%%%%%%%%%%%%%%%%%%%%%%%%%
%
The modified BS rule gets the form
\begin{equation}
\label{sec:2 gBS}
\int_{x_0}^{x_1} \sqrt{E_n - V(x)} \mathrm{d} x 
= \pi \left(n+\frac{1}{2} \right)-\frac{1}{2} \varphi(a) \, .
\end{equation}
The new term $\varphi(a)$ is a quite involved relation of Gamma functions
\begin{equation}
\varphi(a) = a \left(1-\ln(a) \right) + \frac{1}{2i} \ln \left(\frac{\Gamma(1/2+ia)}{\Gamma(1/2-ia) \left[1+\exp(-2\pi a) \right]} \right), 
\end{equation}
where
%
%%%%%%%%%%%%%%%%%%%%%%%%%%%%%%%%%%
\begin{equation}
a = \frac{1}{\pi} \int_{x_1}^{x_2} \left( -p^2 \right)^{1/2} \mathrm{d}x,
\qquad p = \sqrt{E_n-V(x)}.
\end{equation}
%%%%%%%%%%%%%%%%%%%%%%%%%%%%%%%%%%
%
For states which are found in the lower part of the bound region the generalized Bohr-Sommerfeld rule \eqref{sec:2 gBS} can be simplified as follows  \cite{2013waap.book.....K}
%
%%%%%%%%%%%%%%%%%%%%%%%%%%%%%%%%%%
\begin{equation}
\int_{x_0}^{x_1} \sqrt{E_n - V(x)} \mathrm{d} x 
= \pi \left(n+\frac{1}{2} \right)-\frac{i}{4} \exp\left( 2i \int_{x_1}^{x_2} \sqrt{E_n-V(x)} \mathrm{d}x \right),
\label{sec:2 sgBS}
\end{equation}
%
%%%%%%%%%%%%%%%%%%%%%%%%%%%%%%%%%%
%
where $x_2$ is the third classical turning point right of the potential barrier. The additional term introduces an exponentially small imaginary part for $E_n $, which is a measure for the barrier penetrability and depends strongly on the area of the potential barrier above $E_n$, see Fig. \ref{fig:1}. 

Since the imaginary part  of (\ref{sec:2 sgBS}) is small one may approximately treat the integral in the exponential function as real and integrate only along  the real axis. 
By writing the energy as $E_n = E_{0n}+ i E_{1n}$, with real part $E_{0n}$ and small imaginary part $E_{1n}$, one can simplify (\ref{sec:2 sgBS}) even further. Additionally, if $E_{1n} \ll E_{0n}$, one may expand the square root on the left side of (\ref{sec:2 sgBS}) and reduce the problem into the solution of two considerably simpler equations

%
%%%%%%%%%%%%%%%%%%%%%%%%%%%%%%%%%%
\begin{align}
&\int_{x_0(E_{0n})}^{x_1(E_{0n})} \sqrt{E_{0n}-V(x)} \mathrm{d} x
=\pi \left(n+\frac{1}{2} \right) \, ,
\label{constant-NBS}
\\
& E_{1n} = -\frac{1}{2}\exp\left(2 i \int_{x_1(E_{0n})}^{x_2(E_{0n})} \sqrt{E_{0n}-V(x)} \mathrm{d} x \right) \left( \int_{x_0(E_{0n})}^{x_1(E_{0n})} \frac{1}{\sqrt{E_{0n}-V(x)}} \mathrm{d} x \right)^{-1} \, .
\label{constant-E1}
\end{align}
%%%%%%%%%%%%%%%%%%%%%%%%%%%%%%%%%%
%
Here, the turning points are calculated with respect to $E_{0n}$. More details are provided in Sec. \ref{Details for BS Expansion}.
The first of the equations is the classical Bohr-Sommerfeld rule for bound states and fully determines the real part $E_{0n}$ as might be expected. The second equation is in agreement with the well known Gamow formula \cite{2013waap.book.....K}, where $E_{1n}$ is directly calculated by inserting the value of $E_{0n}$ estimated in the previous step. In a sense this is a two step procedure.
%

%%%%%%%%%%%%%%%%%%%%%%%%%%%%%%%%%%
\subsubsection{Higher Order Correction to the Real Part}
%%%%%%%%%%%%%%%%%%%%%%%%%%%%%%%%%%
%
The accuracy in the calculation of the real part of the energy for the bound states can be improved by adding higher order corrections to the classical Bohr-Sommerfeld rule. The extended Bohr-Sommerfeld rule for the bound states with the leading correction term, as shown in \cite{2013waap.book.....K} has the form
%
%%%%%%%%%%%%%%%%%%%%%%%%%%%%%%%%%%
\begin{equation}
\int_{x_0}^{x_1} \sqrt{E_n - V(x)} \mathrm{d} x = \pi \left(n+\frac{1}{2} \right)+\frac{1}{24} \frac{\partial^2}{\partial E_n^2} \int_{x_0}^{x_1} \left(\frac{\mathrm{d} V(x)}{\mathrm{d}x} \right)^2 \frac{\mathrm{d}x}{\sqrt{E_n - V(x)}} \, .
\label{bsho}
\end{equation}
%%%%%%%%%%%%%%%%%%%%%%%%%%%%%%%%%%
%
From the previous expansion of the Bohr-Sommerfeld rule for quasi-stationary states \eqref{constant-NBS} it is known that the real part of the energy of low lying states  can be approximated  using the classical Bohr-Sommerfeld rule. If the potential barrier is large enough, one should expect that the problem for the real part approaches the pure bound state problem, because the imaginary part and tunneling effects become negligible. It is an interesting question whether using the higher order correction for bound states \eqref{bsho} also yields better results for quasi-stationary states. In other words whether a more accurate estimation of $E_{0n}$ will lead to a similar improvement in the accuracy for $E_{1n}$ via Gamow's formula (\ref{constant-E1}). We want to address this educated guess by comparing the results numerically.

%We address this question in two steps. First we use the higher order Bohr-Sommerfeld rule \eqref{bsho} to find the real part, afterwards we determine the imaginary part by using the Gamow formula \eqref{constant-E1}. 
%
The two-step method described earlier can resolve with quite high accuracy the spectrum if the form of the potential allows for analytic evaluation of the  integrals. If instead the potential is complicated there are two ways that one may use in estimating the spectra without relying on full scale numerical techniques. The first approach is based on the numerical evaluation of the integrals and the second in fitting the potentials with simple functions for which the integrals can be evaluated analytically.

%%%%%%%%%%%%%%%%%%%%%%%%%%%%%%%%%%
\subsubsection{Numerical evaluation of the generalised Bohr-Sommerfeld rule}
\label{numerical-BS}
%%%%%%%%%%%%%%%%%%%%%%%%%%%%%%%%%%
%
The non-trivial nature of the potentials can be met in astrophysically relevant problems and call for numerical evaluation of the generalized Bohr-Sommerfeld approach. In other words the numerical evaluation of the relations (\ref{constant-NBS}), (\ref{constant-E1}) and (\ref{bsho}).
Fortunately the numerical tools necessary to solve this set of equations are rather simple to implement and one does not need to solve the ODE \eqref{wave-equation}  numerically as part of a boundary value problem. In this way one avoids several problems related to convergence or integration in the complex plane. Furthermore,  the Bohr-Sommerfeld approach can serve as an independent and simple check for other methods if they are applied to new problems for which no literature values are available.

Both Bohr-Sommerfeld rules \eqref{constant-NBS} and \eqref{bsho} can numerically be solved via root finding and integration. Each of the rules corresponds to a one dimensional function in $E_{0n}$ for a given choice of $n$. The root of each function can be found and corresponds to the correct solution for $E_{0n}$. To determine the imaginary part $E_{1n}$ one inserts the result for $E_{0n}$ in the Gamow formula \eqref{constant-E1}, which can directly be computed. The final result is now simply given by $E_n = E_{0n}+iE_{1n}$.
%
%
%
%%%%%%%%%%%%%%%%%%%%%%%%%%%%%%%%%%
\section{Analytic Fitting of the Potentials }
\label{analytictoymodel}
%%%%%%%%%%%%%%%%%%%%%%%%%%%%%%%%%%
%
When studying a complicated problem it is reasonable to test our methods using approximate problems which, if possible, include all the ingredients of the full problem.
Typically, in the astrophysically relevant cases the potentials are quite complicated and lead to integrals that are intractable with purely analytical methods.  However, analytical results, even if approximate, are always desirable, since one may draw  general conclusions without relying on numerical calculations. 

Our approach will be based on fitting the potential with simpler functions, which can be treated analytically. This was the approach followed in earlier works related to the study of quasi-normal model of black holes, see for example \cite{1984PhLA..100..231B, 1984PhRvL..52.1361F, 1985ApJ...291L..33S}.  The simplest and most known result of this approach was the formula derived by Schutz and Will \cite{1985ApJ...291L..33S}, where the true potential was approximated by a parabola. This formula proved  to be a good approximation for the first quasi-normal modes of Schwarzschild black holes %with mass $M$
%
%%%%%%%%%%%%%%%%%%%%%%%%%%%%%%%%%%
\begin{equation}
\left( M \omega_n \right)^2 = V_\text{max} - i \left(n+\frac{1}{2}\right) \sqrt{-2 V^{\prime \prime}_\text{max}} \, ,
\end{equation}
%%%%%%%%%%%%%%%%%%%%%%%%%%%%%%%%%%
%
where $V_\text{max}$ is the value at the maximum of the Regge-Wheeler potential barrier and $V^{\prime \prime}_\text{max}$ its second derivative at the same point. 
The toy model used here to treat the perturbations of ultra-compact stars is designed in the same spirit.

As it will be shown in the next section, it turns out that the most appropriate way for treating the perturbations of ultra-compact stars is to use two fitting functions. One for the bound region and one for the potential barrier. Here we approximate the bound region with the quartic oscillator $U_\text{Q}$ and the potential barrier with the Breit-Wigner distribution function $U_\text{BW}$ defined as
%
%%%%%%%%%%%%%%%%%%%%%%%%%%%%%%%%%%
\begin{equation}
U_\text{Q}(x) = U_0 + \lambda_{0}^2(x-x_\text{min})^4 \, ,  \quad \mbox{and} \quad U_\text{BW}(x) = \frac{U_{1}}{1+\lambda_1(x-x_\text{max})^2} \, .
\label{approximated_potentials}
\end{equation}
%%%%%%%%%%%%%%%%%%%%%%%%%%%%%%%%%%
%
The two functions depend altogether on six parameters ($x_\text{min}, x_\text{max}, U_0, U_1, \lambda_0, \lambda_1$). Still, the final result only depends on four of them ($U_0, U_1, \lambda_0, \lambda_1$). The values of these parameters are fixed by matching them with the true potentials. The parameters $U_0$ and  $U_1$ will be the interior minimum and maximum 
%$(V_\text{min}, V_\text{max})$ 
of the true potential. Identifying the second derivative $V^{\prime \prime}_\text{max}$ at the barrier maximum with the second derivative of the Breit-Wigner distribution function yields $\lambda_1 = -  V^{\prime\prime}_\text{max}/2 V_\text{max}$. The calculation of the remaining parameter $\lambda_0$ is not  straight forward because the second derivative at the minimum does not yield an optimal overall fit of the entire bound region. 
Thus, for constant density stars we demand the quartic oscillator to be equal to the Regge-Wheeler potential at the surface of the star. However, for gravastars we keep $U_0=V_\text{min}$ but define $x_\text{min}$, to be in the middle of the two $x$-values satisfying $V(x)=U_1$. Finally, $\lambda_0$ follows from the demand that $U_\text{Q}(x_\text{max})=V_\text{max}$. 
To get the solution for $E_n= E_{0n}+i E_{1n}$ we first use the Bohr-Sommerfeld rule \eqref{constant-NBS} to calculate $E_{0n}$ of the quartic oscillator. This is a straightforward procedure and one finds the following analytic expression
\begin{equation}
E_{0n} =  U_0 + \lambda_0^{2/3}\left(  \frac{3 \pi }{2 \sqrt{2} \text{EllipticK}(1/\sqrt{2})}  \left(n+\frac{1}{2} \right)  \right)^{4/3} \, ,
\label{quartic-low}
\end{equation}
where $\text{EllipticK}(x)$ is the complete elliptic integral of the first kind. 
Next we use the Breit-Wigner distribution function and the quartic oscillator described by \eqref{approximated_potentials} in the Gamow  formula \eqref{constant-E1} in order to determine $E_{1n}$. This leads to the following analytic expression:
%
%%%%%%%%%%%%%%%%%%%%%%%%%%%%%%%%%%
\begin{align}
E_{1n}
=&-\frac{1}{4} \sqrt{\lambda_0}\left(E_{0n}-U_{0} \right)^{1/4} \frac{\Gamma(3/4)}{\sqrt{\pi}\Gamma(5/4)} 
\nonumber \\
&\times \exp\left(4i  \sqrt{\frac{U_{1}-E_{0n}}{\lambda_1}} \text{EllipticE}\left(i \sqrt{\frac{U_{1}}{E_{0n}}-1}, i \frac{\sqrt{E_{0n}/U_{1}}}{\sqrt{1-E_{0n}/U_{1}}} \right) \right) \, ,
\label{BW-low}
\end{align}
%%%%%%%%%%%%%%%%%%%%%%%%%%%%%%%%%%
%
where $\text{EllipticE}(x,y)$ is the elliptic integral of the second kind and $\Gamma(x)$ is the gamma function. More details are given in Sec. \ref{Details for the Analytic Model}.

%Having the real and imaginary part we now found the solution for $E_n$. 
The corresponding quasi-normal modes $\omega_n$ are finally derived by inserting \eqref{quartic-low} and \eqref{BW-low} in
\begin{align}
\omega_n = \sqrt{E_n} = \sqrt{E_{0n}+iE_{1n}} \, .
\end{align}
This is the fully analytic approximate result for all trapped states of ultra-compact stars and gravastars and will be applied in the calculation of their spectra in the next section.
%
%
%
%
%%
%%%%%%%%%%%%%%%%%%%%%%%%%%%%%%%%%%

%%%%%%%%%%%%%%%%%%%%%%%%%%%%%%%%%%
\section{Applications}
\label{applications}
%%%%%%%%%%%%%%%%%%%%%%%%%%%%%%%%%%
%
In this section we demonstrate the efficiency of the Bohr-Sommerfeld methods by applying them to the perturbations of two physical systems, the well known constant density stars and the more exotic gravastars, assuming no-rotation. In both cases the axial gravitational  perturbations reduce to  one dimensional wave equations of the form
%
%%%%%%%%%%%%%%%%%%%%%%%%%%%%%%%%%%
\begin{equation}
\frac{\mathrm{d}^2}{\mathrm{d} {r^{*}}^2} \Psi(r) + \left(\omega_n^2-V(r) \right) \Psi(r) = 0 ,
\end{equation}
%%%%%%%%%%%%%%%%%%%%%%%%%%%%%%%%%%
%
where $r^{*}$ is the so-called tortoise coordinate defined as
%
%%%%%%%%%%%%%%%%%%%%%%%%%%%%%%%%%%
\begin{equation}
r^{*} = r + 2M \ln\left(\frac{r}{2M}-1 \right)  .
\end{equation}
%%%%%%%%%%%%%%%%%%%%%%%%%%%%%%%%%%
For potentials of the form shown in Fig 	\ref{fig:1} one can now apply the Bohr-Sommerfeld rules discussed in the previous section. This type of potentials are met in the perturbations of  ultra-compact objects with radius smaller than the corresponding maximum $(r\approx 3M)$ of the Regge-Wheeler potential 
%
%%%%%%%%%%%%%%%%%%%%%%%%%%%%%%%%%%
\begin{equation}
\label{Regge-Wheeler}
V(r) = \left(1-\frac{2M}{r} \right) \left[ \frac{l(l+1)}{r^2}-\frac{6M}{r^3}\right] \, .
\end{equation}
%%%%%%%%%%%%%%%%%%%%%%%%%%%%%%%%%%
%
This potential characterises the perturbations of Schwarzchild black-holes and in general the perturbations of the exterior spacetime of spherically symmetric bodies.
%

%
%%%%%%%%%%%%%%%%%%%%%%%%%%%%%%%%%%
\begin{figure}
\centering
\includegraphics[width=8cm]{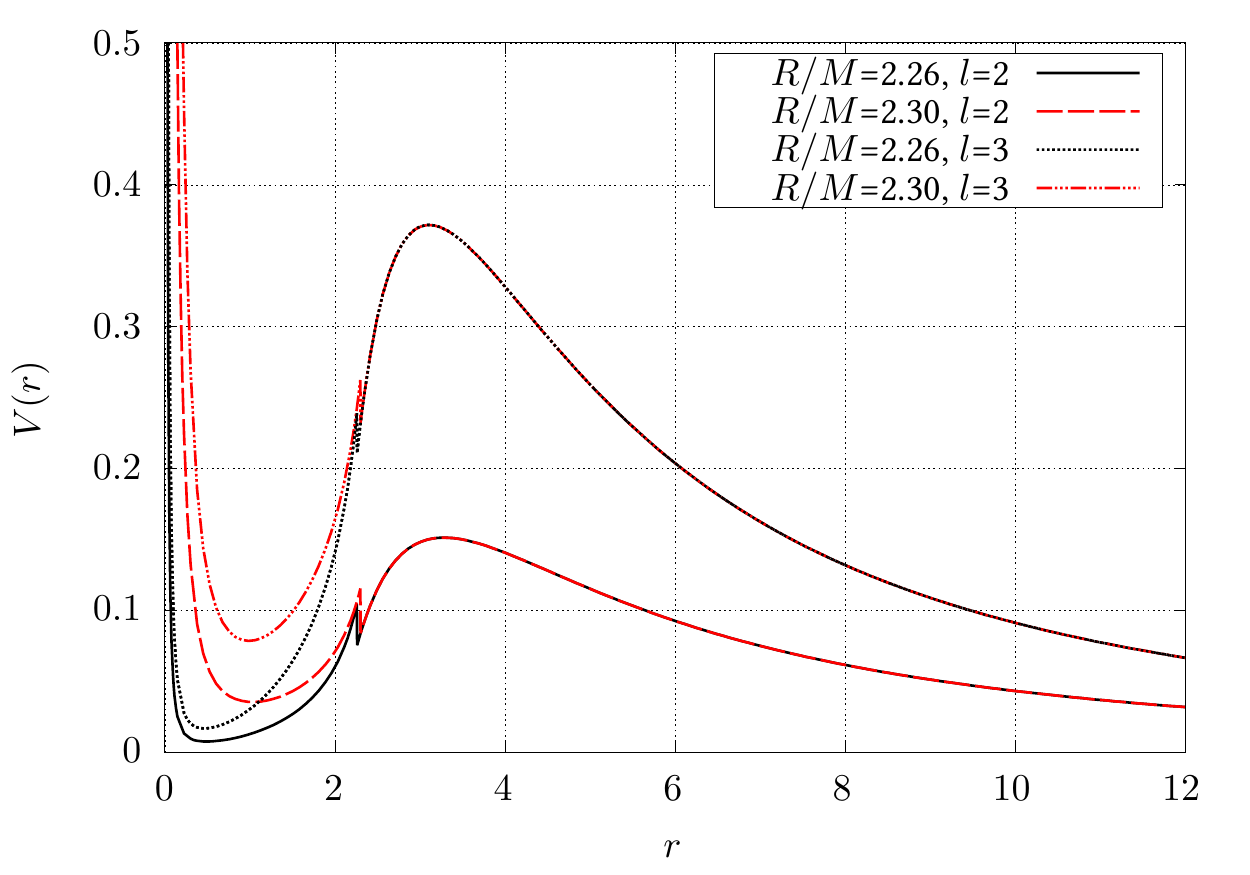}
\includegraphics[width=8cm]{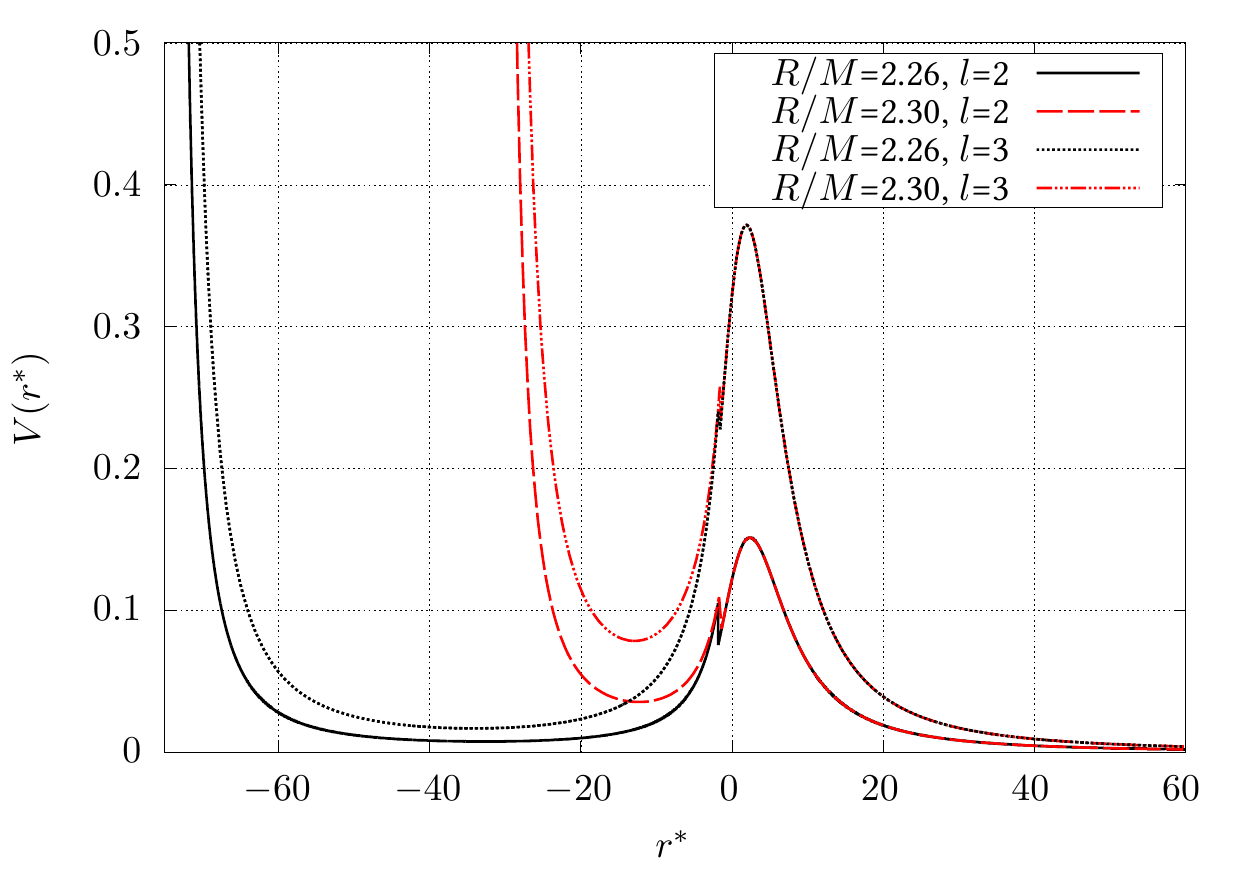}
\caption{The effective potential for constant density stars for different $R/M$ and $l$. Left: with respect to the $r$ coordinate. Right: with respect to the tortoise coordinate $r^{*}$}
\label{cs-potential}
\end{figure}
%
%%%%%%%%%%%%%%%%%%%%%%%%%%%%%%%%%%

%%%%%%%%%%%%%%%%%%%%%%%%%%%%%%%%%%
\subsection{Constant Density Stars}
%%%%%%%%%%%%%%%%%%%%%%%%%%%%%%%%%%
%
Constant density stars have been used extensively in the study of compact object dynamics, because their simplicity allows the derivation of analytic or semi-analytic qualitatively correct results. Their oscillations were initially  studied in \cite{1991RSPSA.434..449C, 1994MNRAS.268.1015K, PhysRevD.60.024004, 2000PhRvD..62j7504F}, while they have been the focus of many more recent efforts \cite{2014PhRvD..90d4069C,Pei2015}. The effective potential inside a constant density star has the form
\begin{align}\label{eq-cs-potential}
V(r) = \frac{e^{2 \nu}}{r^3} \left[l(l+1)r+r^3 (\rho-p(r))-6 M(r) \right] \, ,
\end{align}
where $\rho$ is the density of the star, $p(r)$ its pressure and $M(r)$ the mass inside $r$. The tortoise coordinate $r^{*}$ in which the wave equation appears is defined as
%
%%%%%%%%%%%%%%%%%%%%%%%%%%%%%%%%%%
\begin{align}
r^{*} = \int_{0}^{r} e^{-\nu+\mu} \mathrm{d} r \, ,
\end{align}
%%%%%%%%%%%%%%%%%%%%%%%%%%%%%%%%%%
%
where $e^{2\nu}$ and $e^{2\mu}$ are the $g_{00}$ and $g_{11}$ components of the metric tensor $g_{\mu \nu}$. Actually, an explicit analytic expression for $r^*$ can be found in \cite{2000PhRvD..62h4020P}. The potential described by \eqref{eq-cs-potential} is plotted for both coordinates  in Fig. \ref{cs-potential}. From it one can clearly see that the compactness determines how wide the bound region is, while $l$ fixes the height of the barrier.

The earlier studies \cite{1995RSPSA.451..341K, 1996ApJ...462..855A} revealed that the spectrum associated to this potential accommodates three classes of ``spacetime'' or $w$-modes. The ``trapped'' modes \cite{1991RSPSA.434..449C, 1994MNRAS.268.1015K}, associated with the potential well, the typical $w$-modes \cite{1992MNRAS.255..119K} associated with scattering on the top of the potential barrier (similar to the QNM of black holes) and a class of extremely fast damped modes the so-called $w_{II}$ or ``scattering'' modes \cite{1993PhRvD..48.3467L}. The ones that will be mostly excited in such an ultra-compact object will be the ``trapped'' modes and the fundamental $w$-mode which resembles to the fundamental QNM of a black-hole \cite{2000PhRvD..62j7504F}, see also a detailed discussion in \cite{2016PhRvD..94h4031C}.
%
%\FloatBarrier
%%%%%%%%%%%%%%%%%%%%%%%%%%%%%%%%%%
\subsubsection{Results}\label{cs-results}
%%%%%%%%%%%%%%%%%%%%%%%%%%%%%%%%%%
%
Here we present the results for the constant density stars derived by using fittings with the approximate potentials (\ref{approximated_potentials}) and numerical evaluation of the generalized Bohr-Sommerfeld rule, as described in \ref{numerical-BS}.

% and the direct  numerical solution of the eigenvalue problem from \cite{1994MNRAS.268.1015K}. 

In Table \ref{constant-NBS-table-226-l3} the results for $R/M=2.26$ and $l=2$ and 3 are presented, while more tables can be found in the Appendix \ref{More Results}. The first column, named ``BS Fitting'', shows the analytic solution (fitting of the potentials) discussed in section \ref{analytictoymodel}. The second column, named ``BS Low'', shows the low order Bohr-Sommerfeld result (combination of equations (\ref{constant-NBS}) and (\ref{constant-E1}) ), while the third column, named ``BS High'', includes the next correction term described by equation (\ref{bsho}). The fourth column named ``Numerical'' shows the full numerical result based on the code used in \cite{1994MNRAS.268.1015K}. 

In all cases we observe that the real part for the mode is calculated with higher accuracy than the corresponding imaginary one. This should not be surprising because the real part grows qualitatively linear over a narrow range, while the imaginary part ranges over many orders of magnitudes. 

The toy model performs very good for all $n$ and agrees with the full numerical result typically within a few percent in the real part. For the imaginary part it gives the correct order of magnitude but overestimates it by a factor of $2\sim 3$ for small $n$ and gets better for large $n$. We also want to mention that it agrees with the total number of trapped modes.

Comparing the numerical solution of the two versions of the Bohr-Sommerfeld rule (``BS Low'' vs ``BS High'')  one finds a significant improvement for small $n$ by using the higher correction term. The precision of the higher correction result, in general,  slowly drops for higher $n$. This can be attributed to the growth of the imaginary part since the potential barrier becomes thinner and the bound state approximation used here is not the correct approximation. The standard Bohr-Sommerfeld method slightly improves with rising $n$ as expected, but interestingly still works very good close to the maximum of the barrier.
%
%%%%%%%%%%%%%%%%%%%%%%%%%%%%%%%%%%
\begin{table}[h!]
\centering
\begin{tabular}{|c|ll|ll|ll|ll|}
\hline
 \multicolumn{9}{|c|}{$l=2$} \\
\hline
\multirow{2}{*}{n} &
	\multicolumn{2}{c|}{\text{BS Fitting}} &
	\multicolumn{2}{c|}{\text{BS Low}} &
	\multicolumn{2}{c|}{\text{BS High}} &
	\multicolumn{2}{c|}{\text{Numerical}} \\
& \text{Re}($\omega_n$) & $|$\text{Im}($\omega_n$)$|$ 
& \text{Re}($\omega_n$) &$|$\text{Im}($\omega_n$)$|$ 
& \text{Re}($\omega_n$) & $|$\text{Im}($\omega_n$)$|$
& \text{Re}($\omega_n$) & $|$\text{Im}($\omega_n$)$|$ \\
\hline
0	&	0.1060	&	2.88e-09	&	0.1068	&	1.01e-09	&	0.1091	&	1.29e-09	&	0.1090	&	1.24e-09	\\
1	&	0.1530	&	9.54e-08	&	0.1462	&	3.16e-08	&	0.1485	&	3.73e-08	&	0.1484	&	3.95e-08	\\
2	&	0.1970	&	1.14e-06	&	0.1856	&	4.13e-07	&	0.1879	&	4.72e-07	&	0.1876	&	5.47e-07	\\
3	&	0.2377	&	7.92e-06	&	0.2251	&	3.47e-06	&	0.2274	&	3.88e-06	&	0.2267	&	4.85e-06	\\
4	&	0.2756	&	3.97e-05	&	0.2646	&	2.28e-05	&	0.2668	&	2.52e-05	&	0.2654	&	3.23e-05	\\
5	&	0.3114	&	1.60e-04	&	0.3040	&	1.40e-04	&	0.3063	&	1.52e-04	&	0.3036	&	1.72e-04	\\
6	&	0.3453	&	5.54e-04	&	0.3411	&	5.51e-04	&	0.3441	&	6.14e-04	&	0.3410	&	7.30e-04	\\
7	&	0.3778	&	1.71e-03	&	0.3784	&	2.00e-03	&	0.3752	&	1.81e-03	&	0.3777	&	2.30e-03	\\
\hline
\hline
 \multicolumn{9}{|c|}{$l=3$} \\
 \hline
\multirow{2}{*}{n} &
	\multicolumn{2}{c|}{\text{BS Fitting}} &
	\multicolumn{2}{c|}{\text{BS Low}} &
	\multicolumn{2}{c|}{\text{BS High}} &
	\multicolumn{2}{c|}{\text{Numerical}} \\
& \text{Re}($\omega_n$) & $|$\text{Im}($\omega_n$)$|$ 
& \text{Re}($\omega_n$) & $|$\text{Im}($\omega_n$)$|$ 
& \text{Re}($\omega_n$) & $|$\text{Im}($\omega_n$)$|$
& \text{Re}($\omega_n$) & $|$\text{Im}($\omega_n$)$|$ \\
\hline
0	&	0.1474	&	3.46e-13	&	0.1494	&	1.34e-13	&	0.1506	&	1.56e-13	&	0.1508	&	1.52e-13	\\
1	&	0.1949	&	1.99e-11	&	0.1886	&	6.70e-12	&	0.1901	&	7.61e-12	&	0.1901	&	7.76e-12	\\
2	&	0.2422	&	4.62e-10	&	0.2279	&	1.37e-10	&	0.2294	&	1.52e-10	&	0.2293	&	1.62e-10	\\
3	&	0.2869	&	5.82e-09	&	0.2671	&	1.71e-09	&	0.2686	&	1.87e-09	&	0.2686	&	2.08e-09	\\
4	&	0.3292	&	4.88e-08	&	0.3064	&	1.54e-08	&	0.3079	&	1.67e-08	&	0.3078	&	1.93e-08	\\
5	&	0.3694	&	3.09e-07	&	0.3457	&	1.10e-07	&	0.3472	&	1.18e-07	&	0.3469	&	1.42e-07	\\
6	&	0.4078	&	1.59e-06	&	0.3850	&	6.54e-07	&	0.3865	&	6.99e-07	&	0.3860	&	8.79e-07	\\
7	&	0.4447	&	6.98e-06	&	0.4243	&	3.45e-06	&	0.4257	&	3.66e-06	&	0.4249	&	4.72e-06	\\
8	&	0.4802	&	2.71e-05	&	0.4636	&	1.83e-05	&	0.4650	&	1.93e-05	&	0.4636	&	2.25e-05	\\
9	&	0.5146	&	9.54e-05	&	0.5019	&	7.32e-05	&	0.5075	&	8.93e-05	&	0.5019	&	9.64e-05	\\
10	&	0.5479	&	3.09e-04	&	0.5403	&	2.83e-04	&	0.5404	&	2.84e-04	&	0.5397	&	3.63e-04	\\
11	&	0.5803	&	9.36e-04	&	0.5782	&	1.03e-03	&	0.5768	&	9.82e-04	&	0.5768	&	1.15e-03	\\
\hline
\end{tabular}
\caption{Results for three versions of the Bohr-Sommerfeld method for uniform density stars with {$R/M$=2.26} and  $l=2$ and 3. The results listed in the first column are derived by using fitting to the true potentials, while the second and third column refer to numerical evaluation of the low and higher order BS rule. In the fourth column the  fully numerical results of \cite{1994MNRAS.268.1015K} are listed for comparison. }
\label{constant-NBS-table-226-l3}
\end{table}
%%%%%%%%%%%%%%%%%%%%%%%%%%%%%%%%%%
%
%\FloatBarrier
%
%
%%%%%%%%%%%%%%%%%%%%%%%%%%%%%%%%%%
\subsection{Gravastars}
\begin{figure}
\centering
\includegraphics[width=8cm]{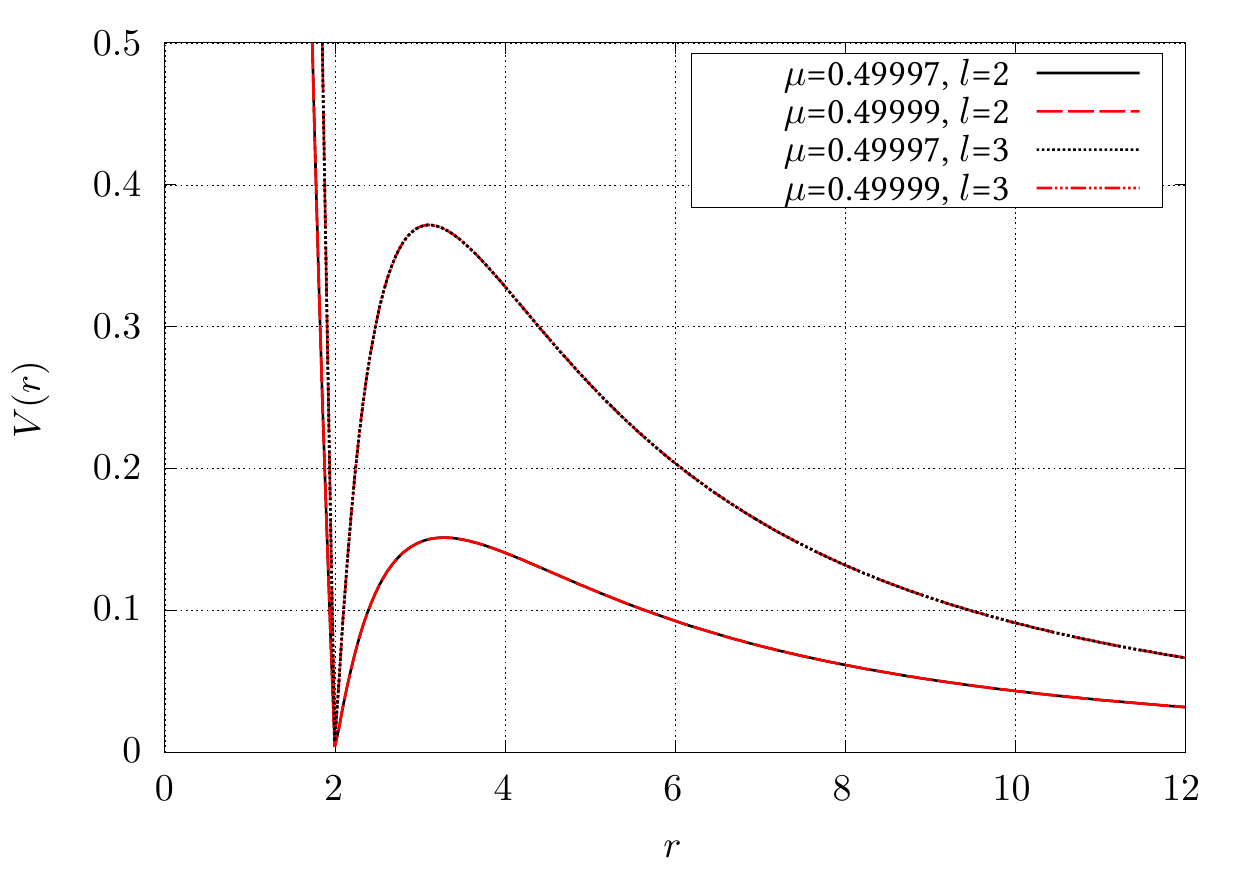}
\includegraphics[width=8cm]{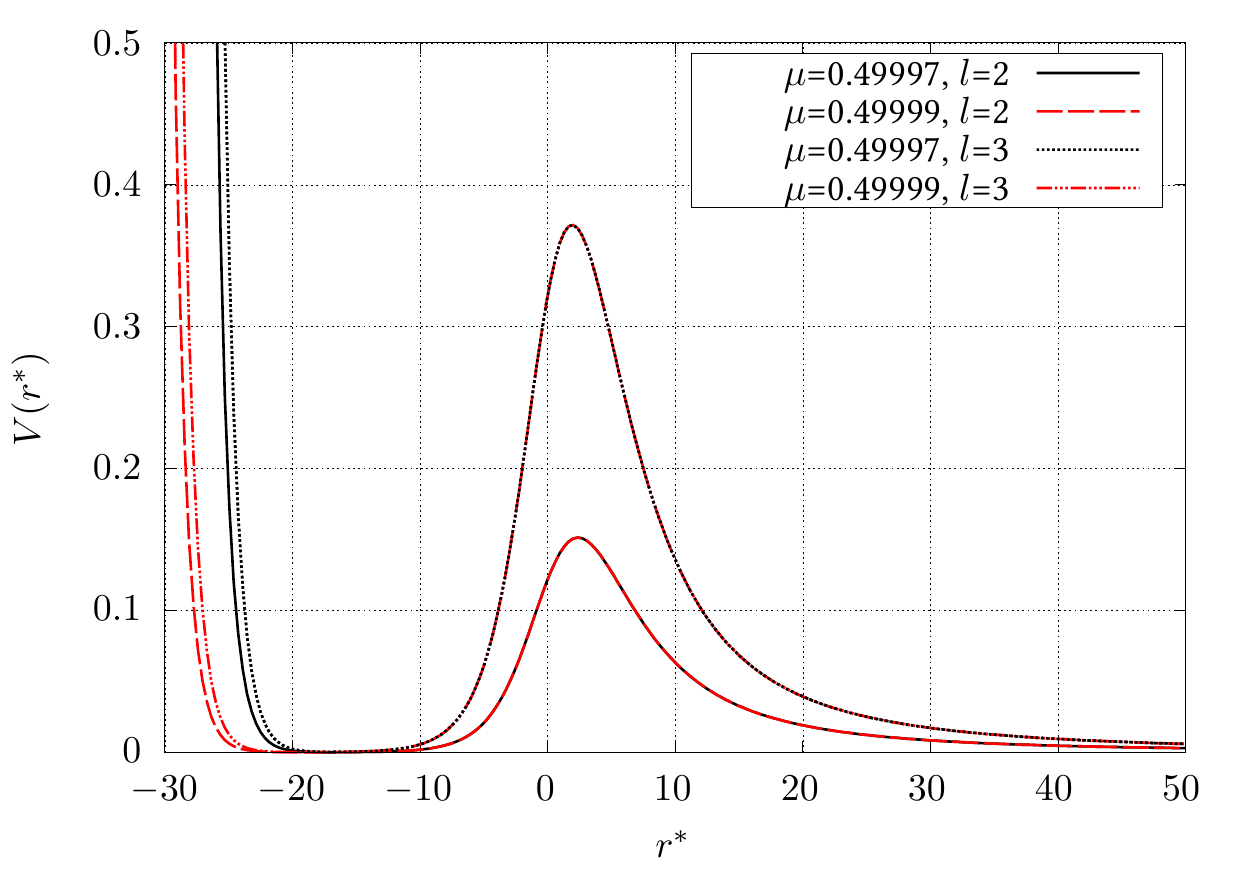}
\caption{Effective potential for gravastars for different values of $\mu=M/a$ and for $l=2$ and 3.  Left: with respect to the $r$ coordinate. Right: with respect to the tortoise coordinate $r^{*}$.}\label{gs-potential}
\end{figure}

%%%%%%%%%%%%%%%%%%%%%%%%%%%%%%%%%%
%
Gravastars are exotic astrophysical objects  proposed as alternative to black holes \cite{2001gr.qc.....9035M}. Depending on the model of interest they are made up of different layers and the interior part consists of a de Sitter condensate with $p=-\rho$. In contrast to constant density stars, whose only parameters are compactness and density, gravastars refer to a larger class of exotic stellar models. In this work we limit ourselves to the thin shell model, since the primary interest is in demonstrating the efficiency of the Bohr-Sommerfeld rule. Here one assumes an infinitely thin shell at radius $a$ which separates the interior and exterior spacetime. More comprehensive work about the physics of gravastars and their perturbations can be found in \cite{2001gr.qc.....9035M, 2004CQGra..21.1135V, 2007CQGra..24.4191C, 2009PhRvD..80l4047P, 1742-6596-222-1-012032, 2014PhRvD..90d4069C, 2016PhRvD..94h4016C}. In \cite{2014PhRvD..90d4069C} a related and interesting WKB analysis is discussed for constant density stars and gravastars in the Eikonal approximation for the $n=0$ fundamental mode. The WKB method \cite{PhysRevA.38.1747} they use differs slightly from the one we use \cite{1991PhLA..157..185P, 2013waap.book.....K}, because the underlying derivation is different. Their imaginary part differs in the integral 
\begin{align}
\int_{x_0}^{x_1} \frac{1}{\sqrt{E_{0n}-V(x)}} \mathrm{d}x \rightarrow 
2 \int_{x_0}^{x_1}\frac{\cos^2\chi(x)}{\sqrt{E_{0n}-V(x)}} \mathrm{d}x,
\end{align}
with $\chi(x) = -\pi/4+\int_{x_0}^{x} \sqrt{E_{0n}-V(x)} \mathrm{d}x$. For rising $n$ they agree and the deviation for small $n$ can be neglected if one compares it with the accuracy of the imaginary part relative to the fully numerical results. Moreover we work without the Eikonal approximation, additionally include the next correction term for the bound state problem and provide results for all trapped states, including small $l \geq 2$.
The interior potential in the model we study is given by
%
%%%%%%%%%%%%%%%%%%%%%%%%%%%%%%%%%%
\begin{align}
\label{eq-gs-potential}
V(r) = \left(1-\frac{8 \pi \rho}{3} r^2\right) \frac{l(l+1)}{r^2}.
\end{align}
%%%%%%%%%%%%%%%%%%%%%%%%%%%%%%%%%%
%
This model corresponds to the specific equation of state in the shell with zero surface density. Since axial perturbations do not couple to matter, our results are also valid for other thin shell models.
Again in the wave equation we use the the tortoise coordinate which inside the star is given by the relation
%
%%%%%%%%%%%%%%%%%%%%%%%%%%%%%%%%%%
\begin{align}
r^* = \sqrt{\frac{3}{8 \pi \rho}} \text{arctanh}\left[ \left(\frac{8 \pi \rho r^2}{3} \right)^{1/2}\right] + C \, .
\label{gravastar-rtortoise}
\end{align}
%%%%%%%%%%%%%%%%%%%%%%%%%%%%%%%%%%
%
Here $C$ is a constant of integration, chosen by demanding that  $r^*$  continuously matched with the usual exterior Schwarzschild tortoise coordinate
\begin{align}
C = a +  2 M \ln\left(\frac{a}{2M}-1 \right) - \sqrt{\frac{3}{8 \pi \rho}} \text{arctanh} \left[ \left(\frac{8 \pi \rho a^2}{3} \right)^{1/2}\right] ,
\end{align}
where $a$ stands for the radius of the gravastar.
The potential \eqref{eq-gs-potential} expressed in both coordinates is shown in Fig. \ref{gs-potential} for different values of the parameter $\mu=M/a$. As for constant density stars, the compactness defines the width of the  the bound region while the angular parameter  $l$  the height of the potential barrier.
%
%%%%%%%%%%%%%%%%%%%%%%%%%%%%%%%%%%
%%%%%%%%%%%%%%%%%%%%%%%%%%%%%%%%%%
%
%
%%%%%%%%%%%%%%%%%%%%%%%%%%%%%%%%%%
\subsubsection{Results}
\label{gs-results}
%%%%%%%%%%%%%%%%%%%%%%%%%%%%%%%%%%
%
The results we find for gravastars share the same properties with  the ultra-compact constant density stars of the previous section \ref{cs-results}. A noticeable difference is that the value of the fundamental  $n=0$ QNM derived via the standard Bohr-Sommerfeld rule is less accurate than the corresponding value for constant density stars. Still when the higher order term, given by equation (\ref{bsho}), is included the agreement with the numerical results is again very good.
\par
Table \ref{gs-NBS-table-0.499970} shows our results for $\mu = 0.49997$ and $l=2$ and 3 while another table for $\mu = 0.49999$ can be found in Sec. \ref{More Results}. The full numerical results are again produced via a code based on the one described in \cite{1994MNRAS.268.1015K}. For the smallest imaginary parts at $n=0$ there is limited accuracy, and potential error of the order of 10-20\,\% are possible.
%%%%%%%%%%%%%%%%%%%%%%%%%%%%%%%%%%
\begin{table}[h!]
\centering
\begin{tabular}{|c|ll|ll|ll|ll|}
\hline
 \multicolumn{9}{|c|}{$l=2$} \\
\hline
\multirow{2}{*}{n} &
	\multicolumn{2}{c|}{\text{BS Fitting}} &
	\multicolumn{2}{c|}{\text{BS Low}} &
	\multicolumn{2}{c|}{\text{BS High}} &
	\multicolumn{2}{c|}{\text{Full Num.}} \\
& \text{Re}($\omega_n$) & $|$\text{Im}($\omega_n$)$|$ 
& \text{Re}($\omega_n$) & $|$\text{Im}($\omega_n$)$|$
& \text{Re}($\omega_n$) & $|$\text{Im}($\omega_n$)$|$ 
& \text{Re}($\omega_n$) & $|$\text{Im}($\omega_n$)$|$ \\
\hline	
0	&	0.1195	&	3.04e-08	&	0.1098	&	6.14e-08	&	0.1307	&	2.11e-07	&	0.1312	&	2.39e-07	\\
1	&	0.2482	&	3.60e-05	&	0.2433	&	3.98e-05	&	0.2495	&	5.09e-05	&	0.2500	&	6.89e-05	\\
2	&	0.3488	&	1.77e-03	&	0.3534	&	2.26e-03	&	0.3520	&	2.15e-03	&	0.3527	&	2.63e-03	\\
\hline
\hline
 \multicolumn{9}{|c|}{$l=3$} \\
 \hline
\multirow{2}{*}{n} &
	\multicolumn{2}{c|}{\text{BS Fitting}} &
	\multicolumn{2}{c|}{\text{BS Low}} &
	\multicolumn{2}{c|}{\text{BS High}} &
	\multicolumn{2}{c|}{\text{Full Num.}} \\
& \text{Re}($\omega_n$) & $|$\text{Im}($\omega_n$)$|$ 
& \text{Re}($\omega_n$) & $|$\text{Im}($\omega_n$)$|$
& \text{Re}($\omega_n$) & $|$\text{Im}($\omega_n$)$|$ 
& \text{Re}($\omega_n$) & $|$\text{Im}($\omega_n$)$|$ \\
\hline
0	&	0.1400	&	7.05e-13	&	0.1277	&	8.42e-12	&	0.1491	&	3.62e-11	&	0.1500	&	4.44e-11	\\
1	&	0.2903	&	2.08e-08	&	0.2763	&	3.07e-08	&	0.2841	&	4.37e-08	&	0.2848	&	5.29e-08	\\
2	&	0.4080	&	4.62e-06	&	0.4021	&	5.47e-06	&	0.4058	&	6.29e-06	&	0.4059	&	7.79e-06	\\
3	&	0.5105	&	2.36e-04	&	0.5145	&	3.16e-04	&	0.5151	&	3.23e-04	&	0.5149	&	3.90e-04	\\
\hline
\end{tabular}
\caption{Results for three versions of the Bohr-Sommerfeld method for gravastars with $\mu$ =0.49997 and for  $l=2$ and 3. The results listed in the first column are derived by using fitting to the true potentials, while the second and third column refer to numerical evaluation of the low and higher order BS rule. In the fourth column we show the fully numerical results of \cite{1994MNRAS.268.1015K}. }
\label{gs-NBS-table-0.499970}
\end{table}
%%%%%%%%%%%%%%%%%%%%%%%%%%%%%%%%%%
%
%%%%%%%%%%%%%%%%%%%%%%%%%%%%%%%%%%
\section{Discussion and Conlusion}
\label{Discussion}
%%%%%%%%%%%%%%%%%%%%%%%%%%%%%%%%%%
%
The results listed in tables \ref{constant-NBS-table-226-l3} and \ref{gs-NBS-table-0.499970} and the ones in the Appendix, demonstrate that by using the higher correction term \eqref{bsho} for the treatment of the semi-bound state problem, one finds extremely accurate results for the real part of the ``trapped'' QNMs, both for constant density stars and for gravastars. This rather simple improvement proved to be extremely useful because the classical Bohr-Sommerfeld rule is in general less accurate for modes in the bottom of the potential barrier (small $n$), while its accuracy improves  for larger values of $n$. The splitting of the complex generalized Bohr-Sommerfeld rule \eqref{sec:2 sgBS} into two real equations is also well justified. Since the imaginary part is expected to be a few orders of magnitude smaller than the real part such a two step approach simplifies the procedure. This was also the reason why we focus in using the high order corrections only for the real part. However, the improvement on the real part induced by equation \eqref{bsho} affects the accuracy in the estimation of the imaginary part.

Both versions of the Bohr-Sommerfeld rule perform better the more bound states exist inside the potential, which is the case for large $l$ and high compactness. Comparing the results from constant density stars and gravastars, it is obvious that gravastars have to be much more compact than constant density stars to accommodate a comparable number of trapped modes. Since the height of the potential barrier is provided by the Regge-Wheeler potential, it does not depend on the compactness but only on the value of $l$. It is therefore the same  for both gravastars and constant density stars. This means that the QNM frequency of the highest trapped mode is in general limited by $l$ and not the by the compactness. Hence, the compactness affects only the size of the bound region and therefore the number of states and their spacing.

It is worth pointing out the excellent performance of the pure analytic fitting method presented in section \ref{analytictoymodel}. This method can be applied to any problem admitting a potential of the type drawn in Fig. \ref{fig:1}.  Keeping in mind that it depends on only four simple to estimate parameters and is based on the standard Bohr-Sommerfeld rule it seems surprising accurate if one compares it with the more complicated versions of the BS method or the fully numerical results. Its simplicity makes it attractive for theoretical studies while the error is typically of the order of a few percent for the real part and still very useful for the imaginary part.
%
%%%%%%%%%%%%%%%%%%%%%%%%%%%%%%%%%%
\subsection{Conclusion}
%%%%%%%%%%%%%%%%%%%%%%%%%%%%%%%%%%
%
We have shown that using a higher order correction to the classical Bohr-Sommerfeld rule allows one to find more accurate results for the quasi-normal frequencies of trapped modes from ultra-compact objects, even if the standard Gamow formula is used. We applied the method to constant density stars and gravastars and found very good agreement with full numerical results. Using the higher order correction term there is a significant improvement in the results for small $n$ for which WKB theory and the Bohr-Sommerfeld rules are usually expected to be worst.

The Bohr-Sommerfeld rule is a general method for problems of that kind and easier to implement than many other fully numerical methods since the numerical solution of the boundary value problem reduces to simple integral evaluation. The BS method also provides a framework to study such problems fully analytically by appropriate fittings of the corresponding potentials. We have successfully solved a general toy model to approximate the trapped modes for ultra-compact stars and found good agreement with numerical methods.

%
%
%%%%%%%%%%%%%%%%%%%%%%%%%%%%%%%%%%%%%%%%%%%%%%%%
\acknowledgments
%%%%%%%%%%%%%%%%%%%%%%%%%%%%%%%%%%%%%%%%%%%%%%%%
The authors would like to thank Vitor Cardoso, Paolo Pani and the anonymous referee for their recommendations which improved the final version of the article and Andreas Boden for useful discussions. SV is grateful for the financial support of the Baden-W\"urttemberg Foundation. 

\bibliographystyle{unsrt}
\bibliography{literatur}

\begin{thebibliography}{10}

\bibitem{LIGO1}
B.~P. {Abbott} and {et~al.}
\newblock Observation of gravitational waves from a binary black hole merger.
\newblock {\em Phys. Rev. Lett.}, 116:061102, Feb 2016.

\bibitem{LIGO2}
B.~P. {Abbott} and {et~al.}
\newblock Tests of general relativity with gw150914.
\newblock {\em Phys. Rev. Lett.}, 116:221101, May 2016.

\bibitem{2016PhRvL.116q1101C}
V.~{Cardoso}, E.~{Franzin}, and P.~{Pani}.
\newblock {Is the Gravitational-Wave Ringdown a Probe of the Event Horizon?}
\newblock {\em Physical Review Letters}, 116(17):171101, April 2016.

\bibitem{2016PhLB..756..350K}
R.~{Konoplya} and A.~{Zhidenko}.
\newblock {Detection of gravitational waves from black holes: Is there a window
  for alternative theories?}
\newblock {\em Physics Letters B}, 756:350--353, May 2016.

\bibitem{1957PhRv..108.1063R}
T.~{Regge} and J.~A. {Wheeler}.
\newblock {Stability of a Schwarzschild Singularity}.
\newblock {\em Physical Review}, 108:1063--1069, November 1957.

\bibitem{1970PhRvL..24..737Z}
F.~J. {Zerilli}.
\newblock {Effective Potential for Even-Parity Regge-Wheeler Gravitational
  Perturbation Equations}.
\newblock {\em Physical Review Letters}, 24:737--738, March 1970.

\bibitem{1973ApJ...185..635T}
S.~A. {Teukolsky}.
\newblock {Perturbations of a Rotating Black Hole. I. Fundamental Equations for
  Gravitational, Electromagnetic, and Neutrino-Field Perturbations}.
\newblock {\em \apj}, 185:635--648, October 1973.

\bibitem{1967ApJ...149..591T}
K.~S. {Thorne} and A.~{Campolattaro}.
\newblock {Non-Radial Pulsation of General-Relativistic Stellar Models. I.
  Analytic Analysis for L $\geq$ 2}.
\newblock {\em \apj}, 149:591, September 1967.

\bibitem{1996PhRvL..77.4134A}
N.~{Andersson} and K.~D. {Kokkotas}.
\newblock {Gravitational Waves and Pulsating Stars: What Can We Learn from
  Future Observations?}
\newblock {\em Physical Review Letters}, 77:4134--4137, November 1996.

\bibitem{1992MNRAS.255..119K}
K.~D. {Kokkotas} and B.~F. {Schutz}.
\newblock {W-modes - A new family of normal modes of pulsating relativistic
  stars}.
\newblock {\em \mnras}, 255:119--128, March 1992.

\bibitem{1991RSPSA.432..247C}
S.~{Chandrasekhar} and V.~{Ferrari}.
\newblock {On the non-radial oscillations of a star}.
\newblock {\em Proceedings of the Royal Society of London Series A},
  432:247--279, February 1991.

\bibitem{1991RSPSA.434..449C}
S.~{Chandrasekhar} and V.~{Ferrari}.
\newblock {On the non-radial oscillations of a star. III - A reconsideration of
  the axial modes}.
\newblock {\em Proceedings of the Royal Society of London Series A},
  434:449--457, August 1991.

\bibitem{1994MNRAS.268.1015K}
K.~D. {Kokkotas}.
\newblock {Axial Modes for Relativistic Stars}.
\newblock {\em \mnras}, 268:1015, June 1994.

\bibitem{PhysRevD.60.024004}
Kazuhiro Tominaga, Motoyuki Saijo, and Kei-ichi Maeda.
\newblock Gravitational waves from a test particle scattered by a neutron star:
  Axial mode case.
\newblock {\em Phys. Rev. D}, 60:024004, Jun 1999.

\bibitem{1999LRR.....2....2K}
K.~D. {Kokkotas} and B.~G. {Schmidt}.
\newblock {Quasi-Normal Modes of Stars and Black Holes}.
\newblock {\em Living Reviews in Relativity}, 2:2, December 1999.

\bibitem{1999CQGra..16R.159N}
H.-P. {Nollert}.
\newblock {TOPICAL REVIEW: Quasinormal modes: the characteristic `sound' of
  black holes and neutron stars}.
\newblock {\em Classical and Quantum Gravity}, 16:R159--R216, December 1999.

\bibitem{2009CQGra..26p3001B}
E.~{Berti}, V.~{Cardoso}, and A.~O. {Starinets}.
\newblock {TOPICAL REVIEW: Quasinormal modes of black holes and black branes}.
\newblock {\em Classical and Quantum Gravity}, 26(16):163001, August 2009.

\bibitem{2011RvMP...83..793K}
R.~A. {Konoplya} and A.~{Zhidenko}.
\newblock {Quasinormal modes of black holes: From astrophysics to string
  theory}.
\newblock {\em Reviews of Modern Physics}, 83:793--836, July 2011.

\bibitem{2001gr.qc.....9035M}
P.~O. {Mazur} and E.~{Mottola}.
\newblock {Gravitational Condensate Stars: An Alternative to Black Holes}.
\newblock {\em arXiv:gr-qc/0109035}, February 2002.

\bibitem{1978amms.book.....B}
C.~M. {Bender} and S.~A. {Orszag}.
\newblock {\em {Advanced Mathematical Methods for Scientists and Engineers}}.
\newblock New York: McGraw-Hill, 1978.

\bibitem{1990CQGra...7L..47G}
J.~W. {Guinn}, C.~M. {Will}, Y.~{Kojima}, and B.~F. {Schutz}.
\newblock {LETTER TO THE EDITOR: High-overtone normal modes of Schwarzschild
  black holes}.
\newblock {\em Classical and Quantum Gravity}, 7:L47--L53, February 1990.

\bibitem{PhysRev.41.721}
J.~L. Dunham.
\newblock The energy levels of a rotating vibrator.
\newblock {\em Phys. Rev.}, 41:721--731, Sep 1932.

\bibitem{1991CQGra...8.2217K}
K.~D. {Kokkotas}.
\newblock {Normal modes of the Kerr black hole}.
\newblock {\em Classical and Quantum Gravity}, 8:2217--2224, December 1991.

\bibitem{0264-9381-10-4-010}
N~Andersson, M~E Araujo, and B~F Schutz.
\newblock Generalized bohr-sommerfeld formula for schwarzschild black hole
  normal modes.
\newblock {\em Classical and Quantum Gravity}, 10(4):757, 1993.

\bibitem{Festuccia:2009:1936-6612:221}
G.~{Festuccia} and H.~{Liu}.
\newblock {A Bohr-Sommerfeld Quantization Formula for Quasinormal Frequencies
  of AdS Black Holes}.
\newblock {\em Advanced Science Letters}, 2(2), 2009.

\bibitem{1991PhLA..157..185P}
V.~S. {Popov}, V.~D. {Mur}, and A.~V. {Sergeev}.
\newblock {Quantization rules for quasistationary states}.
\newblock {\em Physics Letters A}, 157:185--191, July 1991.

\bibitem{2013waap.book.....K}
B.~M. {Karnakov} and V.~P. {Krainov}.
\newblock {\em {WKB Approximation in Atomic Physics}}.
\newblock Springer-Verlag Berlin Heidelberg, 2013.

\bibitem{1984PhLA..100..231B}
H.-J. {Blome} and B.~{Mashhoon}.
\newblock {Quasi-normal oscillations of a schwarzschild black hole}.
\newblock {\em Physics Letters A}, 100:231--234, January 1984.

\bibitem{1984PhRvL..52.1361F}
V.~{Ferrari} and B.~{Mashhoon}.
\newblock {Oscillations of a black hole}.
\newblock {\em Physical Review Letters}, 52:1361--1364, April 1984.

\bibitem{1985ApJ...291L..33S}
B.~F. {Schutz} and C.~M. {Will}.
\newblock {Black hole normal modes - A semianalytic approach}.
\newblock {\em \apjl}, 291:L33--L36, April 1985.

\bibitem{2000PhRvD..62j7504F}
V.~{Ferrari} and K.~D. {Kokkotas}.
\newblock {Scattering of particles by neutron stars: Time evolutions for axial
  perturbations}.
\newblock {\em \prd}, 62(10):107504, November 2000.

\bibitem{2014PhRvD..90d4069C}
V.~{Cardoso}, L.~C.~B. {Crispino}, C.~F.~B. {Macedo}, H.~{Okawa}, and
  P.~{Pani}.
\newblock {Light rings as observational evidence for event horizons: Long-lived
  modes, ergoregions and nonlinear instabilities of ultracompact objects}.
\newblock {\em \prd}, 90(4):044069, August 2014.

\bibitem{Pei2015}
Guancheng Pei and Cosimo Bambi.
\newblock Scattering of particles by deformed non-rotating black holes.
\newblock {\em The European Physical Journal C}, 75(11):560, 2015.

\bibitem{2000PhRvD..62h4020P}
V.~{Pavlidou}, K.~{Tassis}, T.~W. {Baumgarte}, and S.~L. {Shapiro}.
\newblock {Radiative falloff in neutron star spacetimes}.
\newblock {\em \prd}, 62(8):084020, October 2000.

\bibitem{1995RSPSA.451..341K}
Y.~{Kojima}, N.~{Andersson}, and K.~D. {Kokkotas}.
\newblock {On the Oscillation Spectra of Ultra Compact Stars}.
\newblock {\em Proceedings of the Royal Society of London Series A},
  451:341--348, November 1995.

\bibitem{1996ApJ...462..855A}
N.~{Andersson}, Y.~{Kojima}, and K.~D. {Kokkotas}.
\newblock {On the Oscillation Spectra of Ultracompact Stars: an Extensive
  Survey of Gravitational-Wave Modes}.
\newblock {\em \apj}, 462:855, May 1996.

\bibitem{1993PhRvD..48.3467L}
M.~{Leins}, H.-P. {Nollert}, and M.~H. {Soffel}.
\newblock {Nonradial oscillations of neutron stars: A new branch of strongly
  damped normal modes}.
\newblock {\em \prd}, 48:3467--3472, October 1993.

\bibitem{2016PhRvD..94h4031C}
V.~{Cardoso}, S.~{Hopper}, C.~F.~B. {Macedo}, C.~{Palenzuela}, and P.~{Pani}.
\newblock {Gravitational-wave signatures of exotic compact objects and of
  quantum corrections at the horizon scale}.
\newblock {\em \prd}, 94(8):084031, October 2016.

\bibitem{2004CQGra..21.1135V}
M.~{Visser} and D.~L. {Wiltshire}.
\newblock {Stable gravastars -- an alternative to black holes?}
\newblock {\em Classical and Quantum Gravity}, 21:1135--1151, February 2004.

\bibitem{2007CQGra..24.4191C}
C.~B.~M.~H. {Chirenti} and L.~{Rezzolla}.
\newblock {How to tell a gravastar from a black hole}.
\newblock {\em Classical and Quantum Gravity}, 24:4191--4206, August 2007.

\bibitem{2009PhRvD..80l4047P}
P.~{Pani}, E.~{Berti}, V.~{Cardoso}, Y.~{Chen}, and R.~{Norte}.
\newblock {Gravitational wave signatures of the absence of an event horizon:
  Nonradial oscillations of a thin-shell gravastar}.
\newblock {\em \prd}, 80(12):124047, December 2009.

\bibitem{1742-6596-222-1-012032}
P.~{Pani}, E.~{Berti}, V.~{Cardoso}, Y.~{Chen}, and R.~{Norte}.
\newblock Gravitational-wave signature of a thin-shell gravastar.
\newblock {\em Journal of Physics: Conference Series}, 222(1):012032, 2010.

\bibitem{2016PhRvD..94h4016C}
C.~{Chirenti} and L.~{Rezzolla}.
\newblock {Did GW150914 produce a rotating gravastar?}
\newblock {\em \prd}, 94(8):084016, October 2016.

\bibitem{PhysRevA.38.1747}
S.~A. Gurvitz.
\newblock Novel approach to tunneling problems.
\newblock {\em Phys. Rev. A}, 38:1747--1759, Aug 1988.

\end{thebibliography}
%

%
%%%%%%%%%%%%%%%%%%%%%%%%%%%%%%%%%%
\section{Appendix}\label{Appendix}
\subsection{Details for the Expansion}\label{Details for BS Expansion}
The expansion of the generalized BS rule \eqref{sec:2 sgBS} for small $E_{1n}$ is straight forward. 
%\par
After inserting $E_{n}=E_{0n}+iE_{1n}$ in the left-hand side of \eqref{sec:2 sgBS} one finds
\begin{align} \label{lhs}
\int_{x_0}^{x_1} \sqrt{E_{0n}+iE_{1n}-V(x)} \text{d}x 
\approx 
\int_{x_0}^{x_1} \sqrt{E_{0n}-V(x)} \text{d}x + \frac{iE_{1n}}{2} \int_{x_0}^{x_1} \frac{1}{\sqrt{E_{0n}-V(x)}} \text{d}x.
\end{align} 
The same is done for the right-hand side
\begin{align}
&\pi \left( n + \frac{1}{2}\right) - \frac{i}{4} \exp\left(2 i \int_{x_1}^{x_2} \sqrt{E_{0n}+iE_{1n}-V(x)} \text{d}x  \right)
\\
\approx
&\pi \left( n + \frac{1}{2}\right) - \frac{i}{4} \exp\left(2 i \left[ \int_{x_1}^{x_2} \sqrt{E_{0n}-V(x)}\text{d}x +\frac{iE_{1n}}{2} \int_{x_1}^{x_2} \frac{1}{\sqrt{E_{0n}-V(x)}}\text{d}x  \right] \right).\label{rhs}
\end{align}
As it is explained in \cite{2013waap.book.....K} in the section after (4.19), one can neglect the dependence of  $E_{1n}$ on the right-hand side because its contribution is exponentially small. With this we find the final result after comparing the real and imaginary parts of \eqref{lhs} and \eqref{rhs}
\begin{align}\label{real}
\int_{x_0}^{x_1} \sqrt{E_{0n}-V(x)} \text{d}x 
&=
\pi \left( n + \frac{1}{2}\right) ,
\end{align}
and 
\begin{align}
E_{1n} 
&=-\frac{1}{2} \exp\left(2 i \int_{x_1}^{x_2} \sqrt{E_{0n}-V(x)}\text{d}x  \right) \left( \int_{x_0}^{x_1} \frac{1}{\sqrt{E_{0n}-V(x)}}\text{d}x \right)^{-1}.\label{imag}
\end{align}
It is used that $\sqrt{E_{0n}-V(x)}$ is real between $x_0$ and $x_1$ and is imaginary between $x_1$ and $x_2$. The turning points are with respect to $E_{0n}$.
\subsection{Details for the Analytic Model}\label{Details for the Analytic Model}
Here we give more details for the integrals appearing in the analytic toy model in Sec. \ref{analytictoymodel}.
\par
We start with the integral from the classical BS rule for the quartic oscillator \eqref{approximated_potentials}
\begin{align}
\pi\left( n +\frac{1}{2}\right) =\int_{x_0}^{x_1} \sqrt{E_{0n}-U_\text{Q}(x)} \mathrm{d} x
=
&\int_{x_0}^{x_1} \sqrt{E_{0n}-U_0- \lambda_0^2 (x-x_\text{min})^4} \mathrm{d} x.
\end{align}
Defining $u=\lambda_0^{1/2}(x-x_\text{min})/(E_{0n}-U_0)^{1/4}$ one finds
\begin{align}
\int_{x_0}^{x_1} \sqrt{E_{0n}-U_\text{Q}(x)} \mathrm{d} x
&=\frac{(E_{0n}-U_0)^{3/4}}{\lambda_0^{1/2}} \int_{u_0=-1}^{u_1=+1} \sqrt{1-u^4} \mathrm{d} u.
%\\
=\frac{(E_{0n}-U_0)^{3/4}}{\lambda_0^{1/2}} \frac{2}{3} \sqrt{2} \text{EllipticK}\left( \frac{1}{\sqrt{2}} \right).
\end{align}
The new limits of integration are $\pm 1$. This follows from the fundamental definition of turning points, which is $E_{0n}-V(x)=0$, for $x=x_0, x_1$ and that $x_0, x_1$ are real. 
Solving for $E_{0n}$ finally yields
\begin{align}
E_{0n} = U_0 + \lambda_0^{2/3} \left(\frac{3 \pi }{2 \sqrt{2} \text{EllipticK}(1/\sqrt{2})}  \left(n+\frac{1}{2} \right) \right)^{4/3}.\label{real2}
\end{align}
\par
Next we calculate the integrals for the imaginary part. First we determine the integral which is with respect to the quartic oscillator between $x_0$ and $x_1$, including the previous results for the classical BS integral. The integral is given by
\begin{align}
\int_{x_0}^{x_1} \frac{1}{\sqrt{E_{0n}-U_\text{Q}(x)} }\mathrm{d} x
&= \frac{1}{\lambda_0^{1/2} (E_{0n}-U_0)^{1/4}} \int_{u_0=-1}^{u_1=+1} \frac{1}{\sqrt{1-u^4}} \text{d}u
%
%\\
=\frac{1}{\lambda_0^{1/2} (E_{0n}-U_0)^{1/4}} \frac{2 \sqrt{\pi} \Gamma(5/4)}{\Gamma(3/4)}.\label{int1}
\end{align}
\par
Second we calculate the integral between $x_1$ and $x_2$, which is with respect to the Breit-Wigner distribution function \eqref{approximated_potentials}
\begin{align}
\int_{x_1}^{x_2} \sqrt{E_{0n}-U_\text{BW}(x)} \text{d}x 
&=\int_{x_1}^{x_2} \sqrt{E_{0n}-\frac{U_1}{1+\lambda_1(x-x_\text{max})^2}} \text{d}x.
\end{align}
Defining $v = \lambda_1^{1/2} (x-x_\text{max})$ and $\tilde{E_n} = E_{0n}/U_1$ it can be written as
\begin{align}
\int_{x_1}^{x_2} \sqrt{E_{0n}-U_\text{BW}(x)} \text{d}x 
=\left( \frac{U_1}{\lambda_1} \right)^{1/2} \int_{v_1}^{v_2} \sqrt{\tilde{E_n}-\frac{1}{1+v^2}} \text{d}v,
\end{align}
with the new limits of integration $v_{1,2} = \mp \sqrt{1/\tilde{E_n}-1}$, which again follow from the definition of turning points. Using $ 0 < \tilde{E_n} < 1$ one finds
\begin{align}
\int_{v_1}^{v_2} \sqrt{\tilde{E_n}-\frac{1}{1+v^2}} \text{d}v
=-i \frac{\sqrt{\frac{\tilde{E}v^2+\tilde{E}-1}{v^2+1}} \sqrt{v^2+1}(\tilde{E}-1)\sqrt{\frac{\tilde{E}v^2+\tilde{E}-1}{\tilde{E}-1}}}{\tilde{E}v^2+\tilde{E}-1} \text{EllipticE}\left(i v, i \sqrt{\frac{\tilde{E}}{1-\tilde{E}}} \right)\bigg|_{v_1}^{v_2}.
\end{align}
The evaluation of the integral follows from simple algebra after inserting the limits of integration and making use of $\text{EllipticE}\left(- x,y \right)=-\text{EllipticE}\left(x,y \right)$ 
\begin{align}
\int_{v_1}^{v_2} \sqrt{\tilde{E_n}-\frac{1}{1+v^2}} \text{d}v
&= 2 \sqrt{1-\tilde{E}}\, \text{EllipticE} \left(i \sqrt{\frac{1}{\tilde{E}}-1}, i \sqrt{\frac{\tilde{E}}{1-\tilde{E}}} \right).
\end{align}
With this one finds for the entire integral
\begin{align}\label{int2}
\int_{x_1}^{x_2} \sqrt{E_{0n}-U_\text{BW}(x)} \text{d}x 
= 2 \sqrt{\frac{U_1-E_{0n}}{\lambda_1}} \text{EllipticE} \left(i \sqrt{\frac{U_1}{E_{0n}}-1}, i \sqrt{\frac{E_{0n}/U_1}{1-E_{0n}/U_1}} \right). %
\end{align}
The final result \eqref{BW-low} for $E_{1n}$ follows from inserting the results for the integrals.
\subsection{More Results}\label{More Results}

%\label{appendix}
%%%%%%%%%%%%%%%%%%%%%%%%%%%%%%%%%%
%
Here we list additional results for less compact uniform density stars and for more compact gravastars.
%
%%%%%%%%%%%%%%%%%%%%%%%%%%%%%%%%%%
\begin{table}[h!]
\centering
\begin{tabular}{|c|ll|ll|ll|ll|}
\hline
 \multicolumn{9}{|c|}{$l=2$} \\
\hline
\multirow{2}{*}{n} &
	\multicolumn{2}{c|}{\text{BS Fitting}} &
	\multicolumn{2}{c|}{\text{BS Low}} &
	\multicolumn{2}{c|}{\text{BS High}} &
	\multicolumn{2}{c|}{\text{Numerical}} \\
& \text{Re}($\omega_n$) & $|$\text{Im}($\omega_n$)$|$ 
& \text{Re}($\omega_n$) & $|$\text{Im}($\omega_n$)$|$
& \text{Re}($\omega_n$) & $|$\text{Im}($\omega_n$)$|$
& \text{Re}($\omega_n$) & $|$\text{Im}($\omega_n$)$|$ \\
\hline
0	&	0.1753	&	5.37e-07	&	0.1821	&	4.22e-07	&	0.1860	&	5.44e-07	&	0.1856	&	6.20e-07	\\
1	&	0.2444	&	1.88e-05	&	0.2497	&	1.79e-05	&	0.2536	&	2.16e-05	&	0.2519	&	2.66e-05	\\
2	&	0.3107	&	2.86e-04	&	0.3173	&	3.81e-04	&	0.3211	&	4.37e-04	&	0.3160	&	4.63e-04	\\
3	&	0.3724	&	2.66e-03	&	0.3776	&	3.18e-03	&	0.3736	&	2.79e-03	&	0.3764	&	3.64e-03	\\
\hline
\hline
 \multicolumn{9}{|c|}{$l=3$} \\
 \hline
\multirow{2}{*}{n} &
	\multicolumn{2}{c|}{\text{BS Fitting}} &
	\multicolumn{2}{c|}{\text{BS Low}} &
	\multicolumn{2}{c|}{\text{BS High}} &
	\multicolumn{2}{c|}{\text{Numerical}} \\
%n	&\text{WKB T.M.}	 				&\text{Num. B.S.}		&\text{Full Num.}			\\
& \text{Re}($\omega_n$) & $|$\text{Im}($\omega_n$)$|$ 
& \text{Re}($\omega_n$) & $|$\text{Im}($\omega_n$)$|$ 
& \text{Re}($\omega_n$) & $|$\text{Im}($\omega_n$)$|$
& \text{Re}($\omega_n$) & $|$\text{Im}($\omega_n$)$|$ \\
%\hline
\hline
0	&	0.2460	&	8.14e-10	&	0.2543	&	7.24e-10	&	0.2569	&	8.71e-10	&	0.2568	&	9.63e-10	\\
1	&	0.3158	&	4.36e-08	&	0.3213	&	4.53e-08	&	0.3239	&	5.19e-08	&	0.3237	&	6.06e-08	\\
2	&	0.3866	&	1.18e-06	&	0.3884	&	1.18e-06	&	0.3909	&	1.33e-06	&	0.3902	&	1.64e-06	\\
3	&	0.4544	&	1.88e-05	&	0.4554	&	1.98e-05	&	0.4579	&	2.20e-05	&	0.4559	&	2.73e-05	\\
4	&	0.5189	&	2.08e-04	&	0.5205	&	2.38e-04	&	0.5252	&	2.81e-04	&	0.5201	&	3.08e-04	\\
5	&	0.5804	&	1.77e-03	&	0.5840	&	2.05e-03	&	0.5813	&	1.88e-03	&	0.5816	&	2.18e-03	\\
\hline
\end{tabular}
\caption{As in Table \ref{constant-NBS-table-226-l3}, results for $R/M=2.28$ and $l=2$ and 3. }
\label{constant-NBS-table-228-l3}
\end{table}
%%%%%%%%%%%%%%%%%%%%%%%%%%%%%%%%%%
%
%%%%%%%%%%%%%%%%%%%%%%%%%%%%%%%%%%
\begin{table}[h!]
\centering
\begin{tabular}{|c|ll|ll|ll|ll|}
\hline
 \multicolumn{9}{|c|}{$l=2$} \\
\hline
\multirow{2}{*}{n} &
	\multicolumn{2}{c|}{\text{BS Fitting}} &
	\multicolumn{2}{c|}{\text{BS Low}} &
	\multicolumn{2}{c|}{\text{BS High}} &
	\multicolumn{2}{c|}{\text{Numerical}} \\
& \text{Re}($\omega_n$) & $|$\text{Im}($\omega_n$)$|$
& \text{Re}($\omega_n$) & $|$\text{Im}($\omega_n$)$|$
& \text{Re}($\omega_n$) & $|$\text{Im}($\omega_n$)$|$ 
& \text{Re}($\omega_n$) & $|$\text{Im}($\omega_n$)$|$ \\
\hline	
0	&	0.2206	&	7.17e-06	&	0.2313	&	7.84e-06	&	0.2363	&	1.03e-05	&	0.2351	&	1.24e-05	\\
1	&	0.3045	&	2.93e-04	&	0.3175	&	4.86e-04	&	0.3224	&	5.77e-04	&	0.3164	&	5.74e-04	\\
\hline
\hline
 \multicolumn{9}{|c|}{$l=3$} \\
 \hline
\multirow{2}{*}{n} &
	\multicolumn{2}{c|}{\text{BS Fitting}} &
	\multicolumn{2}{c|}{\text{BS Low}} &
	\multicolumn{2}{c|}{\text{BS High}} &
	\multicolumn{2}{c|}{\text{Numerical}} \\
& \text{Re}($\omega_n$) & $|$\text{Im}($\omega_n$)$|$
& \text{Re}($\omega_n$) & $|$\text{Im}($\omega_n$)$|$
& \text{Re}($\omega_n$) & $|$\text{Im}($\omega_n$)$|$ 
& \text{Re}($\omega_n$) & $|$\text{Im}($\omega_n$)$|$ \\
\hline
0	&	0.3109	&	3.68e-08	&	0.3227	&	4.45e-08	&	0.3260	&	5.38e-08	&	0.3258	&	6.36e-08	\\
1	&	0.3963	&	2.27e-06	&	0.4080	&	3.26e-06	&	0.4112	&	3.75e-06	&	0.4102	&	4.66e-06	\\
2	&	0.4834	&	7.52e-05	&	0.4932	&	1.14e-04	&	0.4964	&	1.27e-04	&	0.4925	&	1.40e-04	\\
3	&	0.5671	&	1.51e-03	&	0.5731	&	1.80e-03	&	0.5711	&	1.69e-03	&	0.5704	&	2.00e-03	\\
\hline
\end{tabular}
\caption{As in Table \ref{constant-NBS-table-226-l3}, results for $R/M=2.30$ and $l=2$ and 3. }
\label{constant-NBS-table-230-l3}
\end{table}
%%%%%%%%%%%%%%%%%%%%%%%%%%%%%%%%%%
%%
%%%%%%%%%%%%%%%%%%%%%%%%%%%%%%%%%%
\begin{table}[h!]
\centering
\begin{tabular}{|c|ll||ll|ll|ll|}
\hline
 \multicolumn{9}{|c|}{$l=2$} \\
\hline
\multirow{2}{*}{n} &
	\multicolumn{2}{c|}{\text{BS Fitting}} &
	\multicolumn{2}{c|}{\text{BS Low}} &
	\multicolumn{2}{c|}{\text{BS High}} &
	\multicolumn{2}{c|}{\text{Numerical}} \\
& \text{Re}($\omega_n$) & $|$\text{Im}($\omega_n$)$|$
& \text{Re}($\omega_n$) & $|$\text{Im}($\omega_n$)$|$ 
& \text{Re}($\omega_n$) & $|$\text{Im}($\omega_n$)$|$ 
& \text{Re}($\omega_n$) & $|$\text{Im}($\omega_n$)$|$ \\
\hline	
0	&	0.1106	&	1.37e-08	&	0.0942	&	1.90e-08	&	0.1158	&	7.68e-08	&	0.1158	&	8.14e-08	\\
1	&	0.2299	&	1.43e-05	&	0.2157	&	1.12e-05	&	0.2229	&	1.52e-05	&	0.2235	&	2.05e-05	\\
2	&	0.3232	&	6.26e-04	&	0.3188	&	6.19e-04	&	0.3202	&	6.49e-04	&	0.3201	&	8.47e-04	\\
\hline
\hline
 \multicolumn{9}{|c|}{$l=3$} \\
 \hline
\multirow{2}{*}{n} &
	\multicolumn{2}{c|}{\text{BS Fitting}} &
	\multicolumn{2}{c|}{\text{BS Low}} &
	\multicolumn{2}{c|}{\text{BS High}} &
	\multicolumn{2}{c|}{\text{Numerical}} \\
& \text{Re}($\omega_n$) & $|$\text{Im}($\omega_n$)$|$
& \text{Re}($\omega_n$) & $|$\text{Im}($\omega_n$)$|$
& \text{Re}($\omega_n$) & $|$\text{Im}($\omega_n$)$|$ 
& \text{Re}($\omega_n$) & $|$\text{Im}($\omega_n$)$|$ \\
\hline
0	&	0.1293	&	2.19e-13	&	0.1078	&	1.55e-12	&	0.1299	&	8.34e-12	&	0.1304	&	2.03e-11	\\
1	&	0.2687	&	5.81e-09	&	0.2416	&	5.29e-09	&	0.2501	&	7.97e-09	&	0.2508	&	9.54e-09	\\
2	&	0.3777	&	1.15e-06	&	0.3568	&	8.25e-07	&	0.3612	&	9.85e-07	&	0.3614	&	1.21e-06	\\
3	&	0.4726	&	5.22e-05	&	0.4616	&	4.42e-05	&	0.4635	&	4.73e-05	&	0.4635	&	5.88e-05	\\
4	&	0.5588	&	1.15e-03	&	0.5573	&	1.21e-03	&	0.5563	&	1.17e-03	&	0.5562	&	1.33e-03	\\
\hline
\end{tabular}
\caption{As in Table \ref{gs-NBS-table-0.499970}, results for $\mu=0.49999$ and $l=2$ and 3. }
\label{gs-NBS-table-0.499990}
\end{table}
%%%%%%%%%%%%%%%%%%%%%%%%%%%%%%%%%%
\nocite{}

\end{document}